\documentclass{aa501}

\usepackage{epsfig}
\usepackage{flafter}
\begin{document}

\title{    Europium abundances in F and G disk dwarfs
           \thanks{Based on observations carried out at the European
                   Southern Observatory, La Silla, Chile}
           \thanks{Tables 2 and 6 are only available in electronic form
                   at \hfill\break {\bf http://cdsweb.u-strasbg.fr/A+A.htx} }
      }

    \author{Andreas Koch \and Bengt Edvardsson}

    \institute{Uppsala Astronomical Observatory, Box 515,
               SE-751\,20 Uppsala, Sweden}

    \authorrunning{A. Koch and B. Edvardsson}
    \titlerunning{Europium in F and G Disk Dwarfs}
    \offprints{Bengt.Edvardsson@astro.uu.se}

    \date{Received 27 March 2001 / Accepted 15 October 2001}

\abstract{
Europium abundances for 74 F and G dwarf stars of the galactic disk have
been determined from the 4129.7\,\AA\ Eu\,{\sc ii} line.
The stars were selected from the sample of Edvardsson et al. (1993)
and [Eu/Fe] shows a smaller scatter and a slightly weaker trend with [Fe/H]
than found by Woolf et al. (1995).
The data of the two analyses are homogenized and merged.
We also discuss the adopted effective temperature scale.
         }

\maketitle

\section{Introduction}
The study of the evolution of the Galaxy and the sites for the production
of chemical elements requires observational data of the gradual changes in
chemical composition of the interstellar medium
as a function of time and position in the Galaxy.
Slightly evolved, solar-type stars are very useful for this purpose,
see e.g. the review by McWilliam (1997).
Here we study europium, which is a readily observable representative of the
so-called $r$-process elements.
These are heavy elements formed by the $rapid$ capture of neutrons on
seed nuclei (much more frequent than the $\beta$-decay lifetimes).
Proposed sites of the $r$ process are quickly evolving core-collapse SNe
and neutron star-neutron star mergers (see e.g. Thielemann et al. 2001,
and references therein).
These are the sites where neutron density and temperature are thought to be
sufficiently high to maintain $r$ processes.

Edvardsson et al. (1993, hereafter EAGLNT) investigated the abundances of
13 chemical elements in 189 disk dwarf stars with well-determined ages and
galactic orbits.
No $r$-process element was, however, investigated in that study.
The most readily measurable such element in solar-type stars is europium
through the 4129\,\AA\ Eu\,{\sc ii} line.
94\% of the europium in the Sun is thought to have been produced in the
$r$ process.
Woolf et al. (1995, WTL below)
determined abundances of Eu for a northern sub-sample of the EAGLNT stars.
Here their investigation is supplemented by Eu abundances for 74 southern
stars of the EAGLNT sample.

In section 2 we present the observations and data reductions,
section 3 gives details of the abundance analysis, while
section 4 discusses the uncertainties in the results, and motivates our
use of the EAGLNT effective temperature scale.
Finally, sections 5 and 6 give our results and conclusions.

\section{Observations and reductions}

\subsection{The observations}

During 19 usable nights in 1994 and 1995 high-resolution
spectra of 74 galactic disk F and G dwarfs were obtained. The stars were
selected from the programme stars of EAGLNT. They cover a
range in declination of $-65^{\circ}\,\le\,\delta\,\le\,+23^{\circ}$
thus giving a good addition to -- and some overlap with -- the northern sample
of WTL.
The metallicities [Fe/H] range from $-1.06$ to 0.26 and they have a
distribution in mean galactocentric radius
$5.88 \le R_{\rm m} \le 10.10$\,kpc, reflecting different sites of star
formation.
In the first observational period in July 1994 a spectrum of the solar
disk-integrated flux was also taken.
For the analysis, the Eu\,{\sc ii} line at 4129.7\,\AA\ was chosen.
Among the Eu lines in the visible spectral range it is the least
blended, and its strength is favourable for abundance determination.

The observations themselves were carried out at the European South
Observatory (ESO) at La Silla, Chile, by means of the 1.4m Coud\'e
Auxiliary Telescope.
A wavelength region of 36\,\AA\ centered at the Eu\,{\sc ii} line at
4130\,\AA\ was observed using the Coud\'e Echelle Spectrograph, and
the ESO CCD detectors \#30 (in 1994, Ford Aerospace FA 2048 L) and
\#34 (in 1995, Loral Lo 2048).
The dispersion was 0.02\,\AA\,per pixel and the spectral resolution was
measured at about 90\,000.
Depending on weather conditions, the stellar
brightnesses etc. the signal-to-noise ratio was about 50 at worst, and
340 for the best spectra. The average S/N was $\approx$ 200.
Observed magnitudes reached down to V\,$=8\fm 3$.

\subsection{The data reductions}
The raw spectra were reduced using the ESO routines IHAP and
MIDAS. An averaged bias was subtracted from each spectrum (including the
calibration frames). Afterwards these were divided by flat-field frames.
If any hits due to cosmic rays or radioactive decays within the detector
affected the region around the Eu\,{\sc ii} line, these were filtered
out.
The continuum was rectified by carefully choosing several points that
seemed to be free of lines. Division by a cubic spline function resulted
in the final shape of the spectra. Wavelength calibration was then done
with the spectrum of a thorium-argon lamp.
For many stars two or more spectra were added to reach a useful S/N ratio.
No differences in line-widths or any suspicion of background residuals
are found when spectra of the same star obtained in the two separate
observing runs are compared.

\section{Analysis}
To determine the europium abundance, synthesized spectra of the region
around the Eu line were calculated with the model atmospheres that were
already used in EAGLNT and Woolf et al. (1995, WTL).
The parameters for $\alpha$\,Cen\,B (HR\,5460) were derived from the
analyses of Smith et al. (1986) and Neuforge-Verheecke \& Magain (1997):
$T_{\rm eff}=5220$\,K, $\log g=4.47$, $\xi_t=1.0$\,km\,s$^{-1}$.
The metallicity was adopted from $\alpha$\,Cen\,A in EAGLNT.

The first step consisted of fitting a synthetic spectrum to the observed
solar spectrum. A detailed atomic line list requested from the compiled
VALD database (Kupka et al. 1999) was used. To give a better fit to
the lines in the region the $gf$ values of five lines (and their
blends) were changed.
Typical changes were of the order 0.05-0.1 dex.
Table 1 displays the lines in the vicinity of our Eu\,{\sc ii} line.

\begin{table}
\caption{The most dominant lines in the synthesized region of the solar
spectrum}
\begin{tabular}{ccrc}
\hline
Wavelength&Origin&$\log gf$&Equivalent width\\
$[$\AA]&&&[m\AA]\\
\hline
4128.748&Fe\,{\sc ii}&$-$3.830&43.3\\
4129.159&Ti\,{\sc ii}&$-$2.330&25.4\\
4129.166&Ti\,{\sc i}&0.131&32.1\\
4129.196&Cr\,{\sc i}&$-$1.374&10.8\\
4129.220&Fe\,{\sc i}&$-$2.280&35.3\\
4129.461&Fe\,{\sc i}&$-$2.160&42.3\\
4129.530&'Fake' (Fe\,{\sc i})&$-$3.355&10.5\\
4129.725&Eu\,{\sc ii}&0.173&31.7\\
4129.965&'Fake' (Fe\,{\sc i})&$-$3.455&19.5\\
4130.037&Fe\,{\sc i}&$-$4.280&24.5\\
4130.038&Fe\,{\sc i}&$-$2.470&40.4\\
\hline
\end{tabular}
\end{table}

Europium appears in two stable isotopes: $^{151}$Eu and $^{153}$Eu.
The small difference leads to different term energies and thus
to an additional broadening of the line. For spectral syntheses the
isotope ratio was assumed to be equal 50\%/50\%. Other authors report
the solar istope ratio as
$N(^{153}{\rm Eu})/N(^{151}{\rm Eu}) = (52 \pm 6)/(48 \pm 6)$
(Hauge 1972), or 55/45 (Mashonkina \& Gehren 2000)
and e.g. the isotope ratio for Procyon
(HR\,2943, one of our programme stars) as ($35 \pm 15)/(65 \pm 15$)
(Kato 1987). Considering their uncertainties we expect no major
error when assuming the 50/50 ratio.

Each isotopic line splits into 16 components due to
hyperfine structure. We calculated these using data from
Brostr\"om et al. (1995).
The wavelengths, relative $gf$ values and solar equivalent widths
for all 32 components are listed in Table 2 (accessible in electronic form).

To improve the fits further, two 'artificial' lines were added as substitutes
for unknown blends (see Table 1).
The Eu line itself was not affected significantly by this.
{}From this fit a solar europium abundance $A_{\rm Eu}=0.46$ was
determined. This differs from the commonly accepted solar value of
A$_{Eu}=0.51$ (e.g. Anders \& Grevesse 1989).
The reason for
this deviation is our use of a pure theoretical solar model atmosphere
which is consistent with our stellar models, whereas
Anders \& Grevesse used the semi-empirical Holweger-M\"uller (1974) model.
WTL derive from their fitted $\log gf$ values an
even lower europium abundance of $A_{\rm Eu}=0.44$.
This small difference is due to differences in the observations.
To keep the analysis of the stellar spectra strictly differential relative to
the Sun we use our determined value for the rest of this paper.
Any modification of the $gf$ value in the calculations would only lead
to a systematic shift of all the resulting abundances, but would not
change the trends that are to be seen later.
Synthetic stellar spectra in the wavelength region between 4128 and 4131\,\AA\
were finally created using the above-determined $gf$ values.
The observed spectra were shifted
in wavelength to match the wavelengths of the synthetic ones.
This offered the possibility of controling the quality of the fit at three points:
With the help of the synthetic spectra the setting of the continuum was
fine-tuned to the left of the
Fe\,{\sc ii} line at 4128.7\,\AA, and, if necessary, shifted vertically.
The second reference point of the continuum was to the red of
the Fe\,{\sc i} line at 4130.2\,\AA.
Finally the Fe\,{\sc ii} line at 4128.7\,\AA\,\,served to
find the adequate convolution profile.
Gaussian and rotational profiles, representing the effects of macroturbulence,
rotation and the instrumental profile, were thus convolved with the synthetic
spectra until the iron line was well fitted in shape and depth.
This line
gave in general the best fit in this region since it is neither blended
nor affected by atomic splitting.
The 4129.4\,\AA\ Fe\,{\sc i} line seems to contain an unknown blend
in the solar spectrum which shows up also in some of the other stars,
but mediocre fits of this line had no influence on the derived europium
abundances.

Fig. 1 shows some of the observed spectra together with the fitted
synthetic spectra. The highly broadened lines of the hotter or fast
rotating stars gave rise to higher uncertainties on the abundances,
since several lines overlap there and form one broad feature. The wings
widen to such a large extent that the continuum is not reached, which
made its setting more difficult and sensitive to possible inadequacies
in the line data.
Nonetheless the routine of abundance determination was the same as for the
less broadened, clear line spectra.

The result of our analysis are logarithmic europium abundances relative to
hydrogen normalized on the Sun, [Eu/H]$_{\rm{II}}$
\footnote{[Eu/H]$_{\rm{II}}$\,=\,$\log\,\frac{(N_{\rm Eu}/N_{\rm H})_*}
{(N_{\rm Eu}/N_{\rm H})_{\odot}}$, measured from the Eu\,{\sc ii} line.}.

\begin{figure}[ht]
\setlength{\unitlength}{1cm}
\epsfig{file=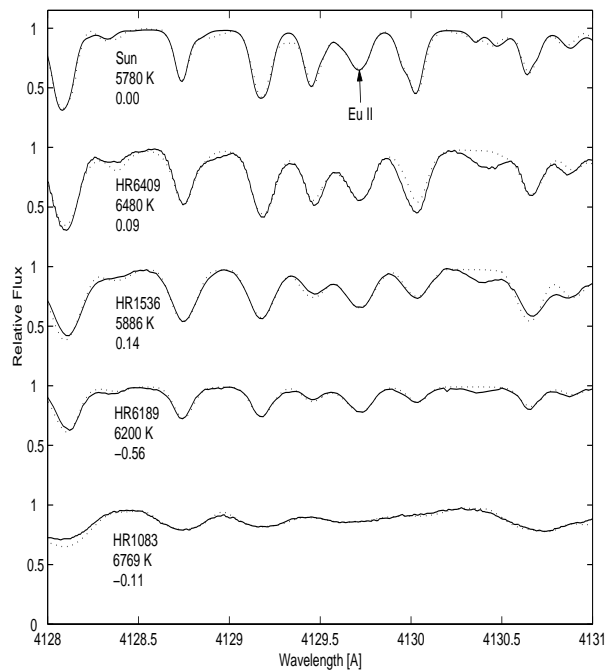,height=9cm,width=8cm,clip=}
\caption{Sample observed spectra (solid lines), the dotted lines show the
synthetic spectra; the feature at 4130\,\AA\, a blend of Gd\,{\sc ii} and
Ni\,{\sc i}, was neglected in the fit.
Given are also the effective temperatures and overall metallicities}
\end{figure}

\section{Error estimates and previous data}

\subsection{Errors in the models}
One error source in these abundance determinations is the
adopted model atmospheres, either by uncertain parameters used there or
by the simplifying assumptions that were made (LTE, plane parallel
atmospheres etc.).
Over the years the MARCS (Gustafsson et al. 1975) model atmospheres were
improved, primarily by updates of the opacities.
The effects of these changes are small: models calculated in the year
2000 give europium abundances lower by 0.01\,dex on average
than the older ones from 1993 used by EAGLNT and WTL.

In general model atmospheres may have errors in the fundamental
parameters. The influence of these error sources on abundances is
discussed at large in EAGLNT.
In that paper it is argued that no uncertainty in temperature larger
than 100\,K should occur, which followed from error estimates of the
basic photometric ($b-y$) data; the analogous estimate for gravity
is 0.2\,dex in $\log g$.

The models and the photometric effective temperature scale derived by EAGLNT
have, however, been challenged.
Blackwell et al. (1995) studied the limb darkening properties of three
solar models: the semiempirical Holweger \& M\"uller (1974, HM) model,
the theoretical flux-constant model of Kurucz (1992, K92) and the theoretical
flux-constant model presented in EAGLNT and used here.
The $T$ vs. $\tau$ relation of the HM model was constructed "by hand" to fit
the profiles of strong spectral lines and the solar limb darkening and
therefore it has a different temperature structure as compared
the flux-constant theoretical models.
It thus fits the observed solar limb darkening quite well, at the cost of not
being flux-constant.
The EAGLNT and Kurucz models are both theoretical, using the mixing-length
approximation for convective energy transport.
The K92 solar model also introduced an "approximate overshooting" scheme on top
of the mixing-length scheme in order to decrease the temperature gradient.
This brought the model to show a limb darkening which is approximately half
way between the HM and the EAGLNT models.
The behaviour of this approximative overshooting recipe for other stars than
the Sun was discussed by Castelli, Gratton \& Kurucz (1997).
They showed that the approximate overshooting scheme can not simultaneously
fit the Sun and other solar-type stars, and this option
has been abandoned in later versions of the K92 model atmospheres programme.
The problem seems to be essentially absent in 3-dimensional hydrodynamic
solar model atmospheres (Asplund et al. 1999).
In our differential analysis the aim is to circumvent systematic errors by
application of the same (non-perfect) analysis to the target stars as to the
reference star, in this case the Sun.

Also the effective temperature scale of EAGLNT has been questioned.
Gratton et al. (1996) used the infrared flux method and interferometric
diameters and the the K92 models with approximate overshooting
to establish an effective temperature scale, which deviates strongly from
the EAGLNT scale as a function of metallicity (see their Fig. 10).
It is not clear to us whether this large systematic difference may be caused
by the use of the approximate overshooting recipe which they later abandoned
(see above).
A metallicity dependent variation of the $T_{\rm eff}$ scale of EAGLNT by
250\,K would have introduced unacceptable deviations from excitation
equilibrium in their sample (EAGLNT Sect. 4.3.4), and very severe
line-to-line scatter in the derived chemical abundances.

Alonso et al. (1994) used $JHK$ photometry to determine effective temperatures
for 550 late-type dwarf and sub-giant stars.
31 of their stars overlap with the sample of EAGLNT, and the mean difference
Alonso et al. minus EAGLNT is $-50$\,K and a scatter of $\pm 80$\,K,
without any obvious trends with $T_{\rm eff}$, $\log g$ or [Fe/H].

Gratton et al. (1996) also used the effective-temperature sensitive H$_\alpha$
line profiles as support for their temperature scale calibration.
Hydrogen line-profile calculations have also been used for $T_{\rm eff}$
calibrations by e.g. Fuhrmann et al. (1993, 1994) and Gardiner et al. (1999).
Barklem et al. (2000, 2001), however, show that the hydrogen line-broadening
theory used in these and previous calculations has substantial systematic
errors which vary with metallicity and effective temperature.
This is due to the neglect of hydrogen self-broadening.
In particular, previous balmer line calibration work has overestimated the
effective temperatures for cool and for metal-poor stars (Barklem et al. 2000).

Gustafsson (1997) also discussed in more detail the reliability and use of
model atmospheres.

These considerations and also the small sensitivity to model parameter
errors of the abundance results discussed next, make us stick to the
$T_{\rm eff}$ scale of EAGLNT.
This also makes our results consistent with our previous data for the same
stars.

To quantify the influence of uncertainties in the model parameters
we carried out alternative model calculations for four representative stars.
Table 3 displays the computed effects on the derived abundances.
The average uncertainty on
europium abundance when using models with
$\Delta$T$_{\rm{eff}}=+100$\,K was $+0.02$\,dex, whereas the models
with a smaller $\log g$ diminished the europium abundances by 0.07
dex on average.

\begin{table}\setcounter{table}{2}
\caption{Effects on [Eu/H]$_{\rm II}$ of model changes for typical programme stars}
\begin{tabular}{lrrr}
\hline
ID & (T$_{\rm{eff}}$, $\log g$, [Fe/H]) &$\Delta T_{\rm eff}$&$\Delta\log g$\\
&&+100K&$-$0.2\,dex\\
\hline
HR\,1083  & (6769, 4.10, $-$0.11)&0.00&$-$0.08\\
HR\,1687  & (6596, 4.15, 0.26)   &0.02&$-$0.07\\
HR\,4903  & (5953, 4.00, 0.24)   &0.03&$-$0.07\\
HD\,199289& (5894, 4.38, $-$1.03)&0.03&$-$0.05\\
\hline
\end{tabular}
\end{table}

The same procedure was carried out in EAGLNT for their error estimates,
and the result for [Fe/H]$_{\rm II}$ was an error of $-0.02$
dex and $-0.09$\,dex for the same changes of
$\Delta T_{\rm eff}$ and $\Delta\log g$, respectively.
If we add our error estimates to those of EAGLNT we get the resulting
uncertainties by model effects in the crucial quantity:
$\Delta$[Eu/Fe]$_{\rm{II}}\approx +0.04$ for temperature changes
and $\Delta$[Eu/Fe]$_{\rm{II}}\approx +0.02$ for gravity changes.
EAGLNT furthermore estimates an uncertainty in the microturbulence
parameter of $\pm 0.3$\,km\,s$^{-1}$.
Such changes in our models gave no significant changes in europium abundances.
Since EAGLNT made sure
that the metallicities adopted for the models are consistent with the
spectroscopic values, they should not cause any discernible error.

Recent NLTE calculations for Eu\,{\sc ii} have been carried out by
Mashonkina \& Gehren (2000).
The slight underpopulation of the ground level and overpopulation of the
excited levels do not contribute significantly to the errors, since most
of the effect cancels out in our differential analysis.

\subsection{Errors in reduction and determination}
Since both continuum setting and the fitting of synthetic spectra to
observations is done via a fit by eye, these may be important error
sources.
We tested the sensitivity to a continuum uncertainty by
introducing a change in the continuum setting of 2\%.
This typically led to abundance changes by 0.02\,dex.
Errors in the iron abundance due to the continuum setting are
found to be negligible in EAGLNT.
The fitting of the synthesized spectra to the observations was found to
be a major error source for some stars. From high S/N spectra of
stars with narrow lines abundances could be derived with an estimated
fitting uncertainty of $\pm 0.01$\,dex.
But in the case of a low S/N ratio or
very broad lines this precision cannot be maintained. Here the fit was
far more difficult and the uncertainties may reach $\pm 0.05$\,dex.

Spectra of the same stars that were observed during different observing
runs and thus differing e.g. in S/N, were also analysed separately.
These were found to give consistent abundances.

Finally all the errors listed here are added in quadrature and we find
$\Delta$[Eu/H]$_{\rm{II}}\approx \pm 0.08$\,dex,
$\Delta$[Fe/H]$_{\rm{II}} \approx \pm 0.11$ (corresponding to the
estimates in EAGLNT), and finally
$\Delta$[Eu/Fe]$_{\rm{II}} \approx \pm 0.05$\,dex for the spectra
of better quality and on the other hand
$\Delta$[Eu/H]$_{\rm{II}}\approx \pm 0.09$\,dex, and
$\Delta$[Eu/Fe]$_{\rm{II}} \approx \pm 0.07$\,dex for the abundances
from broader lines.

\subsection{Previous europium data}
The first larger determination of europium abundances was released 25
years ago by Butcher (1975). In his work he included 32 galactic G
dwarfs. The derived abundances contained uncertainties of about 25\%.
Butchers' [Eu/H] results have a standard deviation of 0.13 from our data.
No systematic shift with metallicity is found when Butchers' data are
compared to ours.

\begin{figure}
\setlength{\unitlength}{1cm}
\epsfig{file=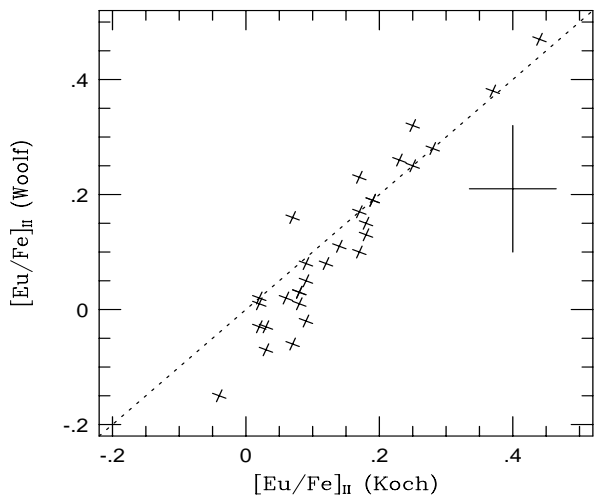,height=7cm,width=8cm,clip=}
\caption{Comparison of Eu abundances with Woolf et al. (1995).
Error bars of the magnitudes estimated in the two investigations are plotted.
The dotted line is unity}
\end{figure}

\begin{table} 
\caption{Comparison of [Eu/Fe]$_{\rm{II}}$ from different projects.
Not all of the WTL stars are included}
\begin{tabular}{lrrrr}
\hline
ID&This &Woolf et al.&Zhao&Butcher\\
&paper&(1995)&(1994)&(1975)\\
\hline
HR\,1101&0.03&$-$0.07& &$-$0.21\\
HR\,2883&0.23&0.26&0.09& \\
HR\,2943&0.02&0.02& &0.04\\
HR\,3018&0.37&0.38&0.32&0.23\\
HR\,3578&0.44&0.47&0.42&0.32\\
HR\,4540&0.02&$-$0.03& &0.12\\
HR\,8181&0.15& &0.11&0.15\\
HD\,215257&0.17& & &0.06\\
\hline
\end{tabular}
\end{table}

One of the programme stars of da Silva et al. (1990) was HR\,3018, which
was also observed by us. They derive an abundance of
[Eu/Fe]$_{\rm{II}}=0.39$, whereas we give its abundance as 0.37.
The deviation is smaller than the estimated uncertainties. In general
HR\,3018 was the star whose abundance coincides in nearly all
observational runs so far. Gratton \& Sneden (1994) report for this star
a value of 0.38, corresponding very well to our result. Their work
contains furthermore our programme star HR\,2883, for which they give an
abundance of $0.14 \pm 0.07$. This is significantly lower than our
value of $0.23 \pm 0.06$, but still consistent within the uncertainties. 
A lower abundance of 0.09 for
this cool metal-poor star was derived by Zhao (1994).
Other stars from Zhao's paper correspond well with our values (see Table 4).
The newer paper of Mashonkina \& Gehren (2000) has 3 stars in common
with our paper.
One of them, HR\,7560, has an abundance of $-0.01$ compared to our derived
[Eu/Fe]$_{\rm{II}}=0.12$.
If the same values of $T_{\rm eff}$ and $\log g$ were used, however,
the abundances would be consistent within the respective uncertainties.

One of the latest and the most extensive studies of europium abundances
in solar type stars was presented by WTL. It
included 81 disk F and G type stars from the EAGLNT sample. They estimated
uncertainties in [Eu/Fe]$_{\rm{II}}$ of about 0.11\,dex. 30 of their
programme stars were also observed by us; the comparison in Fig. 2
shows that our derived abundances are generally slightly higher, and
show a systematic trend with [Eu/Fe].
The error bars are misleading in this plot, however, since the iron abundances,
models and model parameters of EAGLNT were used in both the investigations.
The figure therefore reflects the differences in the line strengths
of the Eu line between the two investigations.
If the slightly different solar Eu abundances were taken into account,
a systematic upward (or leftward) shift of all data points by 0.02\,dex
would occur.
The trend implies that the ratios of our Eu line strengths to those of WTL
increase with increasing metallicity.
Comparisons between reduced spectra from WTL
and the present paper suggest that the cause of this difference is a
larger uncertainty in the continuum level near the Eu line of the WTL
data, due to a considerably shorter observed spectral interval, and
with an Eu line position typically only 3.5\,\AA\ from
the red edge of the spectrum.
This explanation is supported by the larger scatter
in [Eu/Fe] vs. [Fe/H] found for the WTL data in the next section.
Table 4, finally, gives a comparison for some stars analysed by different
authors.

\section{Results and discussion}

Europium abundances of the 74 F and G stars determined during this project
are listed in Table 5 (also available in electronic form).
The remaining parameters are taken from EAGLNT.
Their [Fe/H]$_{\rm II}$ abundance and age for $\alpha$\,Cen\,A (HR\,5459)
are adopted for $\alpha$\,Cen\,B (HR\,5460).

\begin{figure}[ht]
\setlength{\unitlength}{1cm}
\epsfig{file=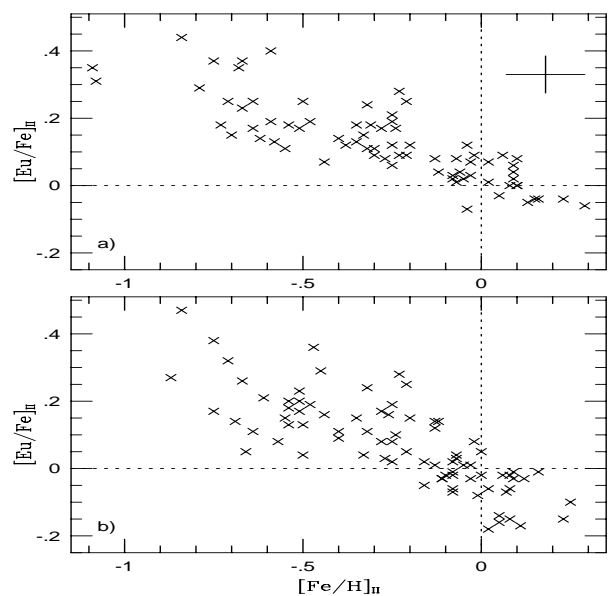,height=8cm,width=8cm,clip=}
\caption{Europium abundance relative to iron vs. iron.
In panel a) our data is shown and the indicated
error bar is of typical size.
In panel b) the corresponding data of Woolf et al. (1995) is shown}
\end{figure}

\begin{table*}[ht]
\caption{Derived Eu abundances. The iron abundances are from EAGLNT}
\begin{tabular}{lrrrr|lrrrr}
\hline
ID &
[Fe/H]$_{\rm I}$&
[Fe/H]$_{\rm II}$&
[Eu/H]$_{\rm II}$&
[Eu/Fe]$_{\rm II}$&
~~~ID &
[Fe/H]$_{\rm I}$&
[Fe/H]$_{\rm II}$&
[Eu/H]$_{\rm II}$&
[Eu/Fe]$_{\rm II}$\\
\hline
HR\,33    &$-$0.38&$-$0.40&$-$0.26&   0.14 &~~~HR\,4657  &$-$0.70&$-$0.71&$-$0.46&   0.25\\
HR\,35    &$-$0.10&$-$0.12&$-$0.08&   0.04 &~~~HR\,4734  &   0.10&   0.10&   0.10&   0.00\\
HR\,107   &$-$0.37&$-$0.35&$-$0.22&   0.13 &~~~HR\,4903  &   0.24&   0.29&   0.23&$-$0.06\\
HR\,140   &   0.05&$-$0.04&$-$0.11&$-$0.07 &~~~HR\,4989  &$-$0.28&$-$0.30&$-$0.19&   0.11\\
HR\,235   &$-$0.15&$-$0.28&$-$0.11&   0.17 &~~~HR\,5338  &$-$0.11&$-$0.07&$-$0.06&   0.01\\
HR\,366   &$-$0.32&$-$0.35&$-$0.17&   0.18 &~~~HR\,5460  &   0.15&   0.19&   0.11&$-$0.08\\
HR\,368   &$-$0.24&$-$0.25&$-$0.06&   0.19 &~~~HR\,5542  &   0.13&   0.09&   0.15&   0.06\\
HR\,370   &   0.12&   0.05&   0.01&$-$0.04 &~~~HR\,5698  &   0.01&   0.08&   0.08&   0.00\\
HR\,573   &$-$0.34&$-$0.30&$-$0.21&   0.09 &~~~HR\,5723  &$-$0.13&$-$0.13&$-$0.05&   0.08\\
HR\,646   &$-$0.32&$-$0.25&$-$0.19&   0.06 &~~~HR\,5996  &   0.23&  0.16 &   0.12&$-$0.04\\
HR\,672   &   0.06&$-$0.07&$-$0.01&   0.08 &~~~HR\,6189  &$-$0.56&$-$0.59&$-$0.40&   0.19\\
HR\,740   &$-$0.25&$-$0.27&$-$0.19&   0.08 &~~~HR\,6243  &   0.00&   0.02&   0.03&   0.01\\
HR\,1010  &$-$0.23&$-$0.33&$-$0.18&   0.15 &~~~HR\,6409  &   0.09&   0.09&   0.13&   0.04\\
HR\,1083  &$-$0.11&$-$0.20&$-$0.08&   0.12 &~~~HR\,6569  &$-$0.27&$-$0.23&$-$0.14&   0.09\\
HR\,1101  &$-$0.11&$-$0.08&$-$0.05&   0.03 &~~~HR\,6649  &$-$0.32&$-$0.34&$-$0.21&   0.11\\
HR\,1173  &   0.09&$-$0.02&   0.07&   0.09 &~~~HR\,6907  &   0.13&   0.10&   0.18&   0.08\\
HR\,1257  &   0.02&   0.04&   0.09&   0.07 &~~~HR\,7126  &   0.21&   0.10&   0.10&   0.00\\
HR\,1536  &   0.14&   0.06&   0.15&   0.09 &~~~HR\,7560  &   0.09&$-$0.04&   0.08&   0.12\\
HR\,1545  &$-$0.33&$-$0.51&$-$0.34&   0.17 &~~~HR\,7875  &$-$0.44&$-$0.38&$-$0.26&   0.12\\
HR\,1673  &$-$0.30&$-$0.25&$-$0.13&   0.12 &~~~HR\,8077  &$-$0.07&$-$0.06&$-$0.02&   0.04\\
HR\,1687  &   0.26&   0.23&   0.19&$-$0.04 &~~~HR\,8181  &$-$0.67&$-$0.70&$-$0.55&   0.15\\
HR\,1983  &$-$0.07&$-$0.03&   0.00&   0.03 &~~~HR\,8665  &$-$0.32&$-$0.21&$-$0.12&   0.09\\
HR\,2233  &$-$0.17&$-$0.21&   0.04&   0.25 &~~~HR\,8697  &$-$0.25&$-$0.31&$-$0.13&   0.18\\
HR\,2354  &   0.13&   0.13&   0.08&$-$0.05 &~~~HR\,8969  &$-$0.17&$-$0.23&   0.05&   0.28\\
HR\,2530  &$-$0.43&$-$0.44&$-$0.37&   0.07 &~~~HD\,6434  &$-$0.54&$-$0.59&$-$0.19&   0.40\\
HR\,2548  &$-$0.20&$-$0.25&$-$0.04&   0.21 &~~~HD\,17548 &$-$0.59&$-$0.62&$-$0.48&   0.14\\
HR\,2835  &$-$0.55&$-$0.54&$-$0.36&   0.18 &~~~HD\,25704 &$-$0.85&$-$0.79&$-$0.50&   0.29\\
HR\,2883  &$-$0.75&$-$0.67&$-$0.44&   0.23 &~~~HD\,51929 &$-$0.64&$-$0.68&$-$0.33&   0.35\\
HR\,2906  &$-$0.18&$-$0.05&$-$0.03&   0.02 &~~~HD\,78747 &$-$0.64&$-$0.67&$-$0.30&   0.37\\
HR\,2943  &$-$0.02&$-$0.08&$-$0.06&   0.02 &~~~HD\,130551&$-$0.62&$-$0.58&$-$0.45&   0.13\\
HR\,3018  &$-$0.78&$-$0.75&$-$0.38&   0.37 &~~~HD\,165401&$-$0.47&$-$0.50&$-$0.25&   0.25\\
HR\,3220  &$-$0.26&$-$0.32&$-$0.08&   0.24 &~~~HD\,188815&$-$0.58&$-$0.55&$-$0.44&   0.11\\
HR\,3578  &$-$0.82&$-$0.84&$-$0.40&   0.44 &~~~HD\,199289&$-$1.03&$-$1.08&$-$0.77&   0.31\\
HR\,4039  &$-$0.38&$-$0.48&$-$0.29&   0.19 &~~~HD\,201891&$-$1.06&$-$1.09&$-$0.74&   0.35\\
HR\,4158  &$-$0.24&$-$0.24&$-$0.07&   0.17 &~~~HD\,210752&$-$0.64&$-$0.73&$-$0.55&   0.18\\
HR\,4395  &$-$0.10&$-$0.03&   0.04&   0.07 &~~~HD\,215257&$-$0.65&$-$0.64&$-$0.47&   0.17\\
HR\,4540  &   0.13&   0.09&   0.11&   0.02 &~~~HD\,218504&$-$0.62&$-$0.64&$-$0.39&   0.25\\
\hline
\end{tabular}
\end{table*}

Fig. 3a displays [Eu/Fe]$_{\rm{II}}$ vs. [Fe/H]$_{\rm{II}}$ for our data.
We use the singly ionized state of both elements since these are the dominant
species in the solar type stars and this ratio is only weakly dependent on
surface gravities and possible overionization effects.
A linear least squares fit to the data yields a slope of
$\Delta$[Eu/Fe]$_{\rm{II}}$/$\Delta$[Fe/H]$_{\rm{II}}=-0.31
\pm 0.02$. At solar metallicities there is an offset in y-direction of
$+0.04 \pm 0.01$.
The scatter in [Eu/Fe]$_{\rm II}$ relative to the linear fit to the
diagram is 0.065\,dex (s.d.).
Following WTL
a more quantitative approach to the scatter around the slope was obtained by
means of a Monte Carlo simulation in which we calculated Gaussian
distributed random pairs of iron and europium abundances.
The estimates from Sect. 4 were taken as representative $1\sigma$ random
errors scattering around the theoretical slope of $-0.31$.
The simulated scatter diagram is similar to the observed scatter.
Since the scatter in our new data set is smaller than that of WTL (see
below), this strengthens their conclusion that most of the scatter in the
results can be explained by observational, analytical and systematic errors
rather than by real stellar scatter of abundances.

Fig. 3b shows the corresponding results from WTL.
Here the slope of a linear fit is $-0.39$ and the line passes the solar
metallicity at [Eu/Fe]$_{\rm II} = -0.02$.
The scatter relative to a linear least squares fit to Fig. 3b is
0.082\,dex, and their Eu abundances are lower than ours at the higher
metallicities.

To homogenize and merge the data we derive a simple linear transformation
between the two data sets in Fig. 3.
Motivated by the smaller scatter in Fig. 3a, and the discussion in the end
of Sect. 4.3, we chose to adjust the WTL data to our results and derive
$\Delta$[Eu/H]$ = $[Eu/H]$({\rm Koch}) - $[Eu/H]$({\rm WTL})
= 0.0619 + 0.0833 $[Fe/H]$_{\rm II}$.
For stars observed in both analyses
the weights were taken to be proportional to inverted squares of the standard
deviations from the linear fits to Figs. 3a and b.
This data for the 125 stars is given in Table 6 (available in electronic form),
which gives [Eu/H]$_{\rm II}$, [Fe/H]$_{\rm II}$, [Eu/Fe]$_{\rm II}$
and $\log {\rm age}$ for each star.

A definite trend is also present in the abundance vs. age diagram for the
merged data set, Fig. 4.
[Eu/Fe]$_{\rm{II}}$ increases by a factor of two over the stellar age range,
although both the Eu and the Fe abundances (relative to hydrogen) decrease with
age.
The isochronic ages were adopted from EAGLNT.
Later new age data has been published by Ng \& Bertelli (1998).
If we plot our abundances against these new ages instead of these used in
EAGLNT we find the same trend but with a larger scatter.

\begin{figure}[ht]
\setlength{\unitlength}{1cm}
\epsfig{file=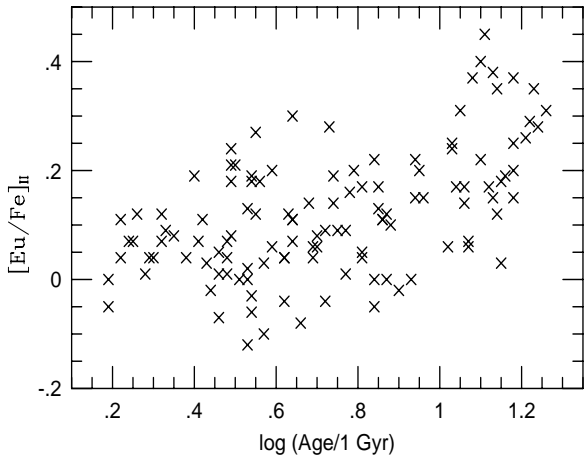,height=7cm,width=8cm,clip=}
\caption{[Eu/Fe] vs. $\log age$ using our data merged with those of
Woolf et al. (1995).  The ages are adopted from EAGLNT}
\end{figure}

Our new results and Eu abundance trends are not greatly different from those
of WTL.
We therefore refer the reader to their discussion of the results.

\section{Conclusions}
The analysis of 74 F and G type galactic disk dwarfs from the sample
of Edvardsson et al. (1993) shows that the abundance ratio [Eu/Fe]$_{\rm{II}}$
decreases with a slope of $-0.31$\,dex/dex with increasing metallicity
in the observed range $-1.06<$\,[Fe/H]\,$<0.29$.
This slope is about 20\% smaller than that found by Woolfe et al. (1995),
and the scatter around the trend is also reduced.

Our data have been merged with those of Woolf et al. (1995), after a slight
adjustment of the latter,
to form a data set of 125 stars with metallicities [Fe/H]$\ga -1.1$.
To improve the data set needed for studies of the formation and evolution of
$r$-process elements, more data for the low and very low metallicity range
is needed.

\begin{acknowledgements}
We would like to thank Vincent Woolf for making available his data and for
swift help.
Bengt Gustafsson is thanked for discussions and suggestions
and the anonymous referee for very helpful comments to the manuscript.
We thank Anders W\"annstr\"om for help and discussions of hyperfine structure
calculations and Sofia Feltzing for performing parts of the observations.
We acknowledge Patrik Thor\'en and
Torgny Karlsson for help with software and discussions.
BE acknowledges support from the
Swedish Natural Science Research Council, NFR.
\end{acknowledgements}

\end{document}